\documentclass[superscriptaddress,showpacs,floatfix,amsmath,amssymb,twocolumn]{revtex4-1}
\usepackage{amssymb}
\usepackage{amsmath}
\usepackage{epsfig}
\usepackage{t1enc}
\usepackage{soul}
\usepackage{color}
\usepackage{times,txfonts}
\newcommand{\beq}{\begin{equation}}
\newcommand{\eeq}{\end{equation}}
\newcommand{\bea}{\begin{eqnarray}}
\newcommand{\eea}{\end{eqnarray}}
\begin{document}
\title{Discriminating quantum field theories in non-inertial frames}
\author{Jason Doukas}
\affiliation{School of Mathematical Sciences, University of Nottingham, Nottingham NG7 2RD, United Kingdom}
\author{Gerardo Adesso}
\affiliation{School of Mathematical Sciences, University of Nottingham, Nottingham NG7 2RD, United Kingdom}
\author{Stefano Pirandola}
\affiliation{Department of Computer Science, University of York, York YO10 5GH, United Kingdom}
\author{Andrzej Dragan}
\address{Institute of Theoretical Physics, University of Warsaw, Ho\.{z}a 69, 00-049 Warsaw, Poland}
\date{January 12, 2015}
\begin{abstract}
Quantum channel discrimination is used to test quantum field theory in non-inertial frames. We search for optimal strategies which can best see the thermality of the Unruh effect.  We find that the usual strategy of counting particles in the vacuum can be improved, thereby enhancing the discrimination. Coherent state probes, which are practical and feasible, give exponential improvement in the discrimination of the Unruh channel and come very close to optimal. In particular, we show that using a short pulse laser, the accelerations required to test the Unruh effect can be reduced by at least three orders of magnitude with the same statistical confidence as could be achieved in vacuum. These results are expected to be relevant to upcoming experimental tests of quantum field theory in curved spacetimes in analogue systems.
\end{abstract}
\maketitle

\section{Introduction}
The Unruh effect \cite{Unruh1976} as it is often understood is the prediction that accelerated bodies in empty space experience a temperature proportional to their acceleration. Despite several experimental proposals being put forth \cite{ExpProp}, verification of the Unruh effect remains an open research program \cite{Doukas2012, Downes2013, Bruschi2010, Alsing2003, Dragan2012, Crispino}. In particular, several experiments are currently underway \cite{Retzker2008,OtherAnalogues} to test analogues of this effect in more accessible regimes.

In 1973 Fulling showed that a quantum field restricted to the region inside the horizon of an accelerated observer (forming a spacetime wedge) could be quantised by performing a generalisation of the canonical quantisation procedure \cite{Fulling1972}. Unruh found in 1976 that by joining two of these Rindler wedges together the Minkowski vacuum state could be written as a product over frequencies of two--mode squeezings between the left-wedge and right-wedge Rindler modes \cite{Unruh1976}.  Rindler modes are a convenient choice of basis because they are either localised in the left or right wedges. Since the accelerated observer has access to only one of these regions, the state available to an accelerated observer can be calculated by writing it in the Rindler basis and tracing over the modes in the other region. This leads to the conclusion that the vacuum appears to an accelerated observer as a thermal state at the Unruh temperature.

The thermal response registered by a uniformly accelerating Unruh-DeWitt detector in vacuum, which can be calculated from the perspective of an inertial frame without reference to the mode decomposition in the accelerated frame, is generally considered to be mathematical confirmation of the Rindler-Fulling quantisation procedure \cite{Unruh1976, Dewitt1979, Birrell1982}. Nevertheless, in scientific enquiry experimental confirmation is always required and it is more strongly demanded the more a theory departs from our ordinary expectations. The prediction that observers in different states of motion disagree on the number of particles is a good example of an occasion in which the theoretical predictions depart strongly from our ordinary expectations. Accepting that the theory should be tested, the question arises as to how one can best do this.

Usually when one thinks about testing the Unruh effect, they think of an accelerated observer detecting particles when there ``should'' have been none. The simplest test one could devise is therefore a test of whether thermal particles are detected (under acceleration) in the vacuum or not. If no particles are detected it would in some sense imply that the vacuum had remained the vacuum (in the sense of being vacuous of particles). We will call this potential situation in which there are no thermal particles observed the \textit{null theory} \cite{footNull}.

While the thermal form of the vacuum state when written in the right wedge Rindler subspace nicely illustrates the physical content of Unruh's result, the effects are not limited to the vacuum state alone. More broadly, the transformation that occurs when changing from the inertial frame to accelerated frame (effectively a change of basis followed by a trace operation), can be thought of as a linear quantum channel \cite{amplificationchannel, Unruh2011}. The terminology \textit{quantum channel} is taken from Shannon's information theory adapted to the quantum setting by quantum information theorists \cite{channeldef}. The Unruh channel is a change of basis that takes any state in the inertial frame to a corresponding state in the accelerated frame.  This opens up the possibility of testing  quantum field theory in curved spacetimes using states other than the vacuum.

Quantum state discrimination has been developed to perform quantum statistical hypothesis testing \cite{QSD}.  For a given input state, the problem of quantum state discrimination is equivalent to quantum channel discrimination: Alice sends a known state to Bob down one of two channels. Bob's task is to identify which of the two channels acted on the received state. The probability of Bob misidentifying the channel can be minimised provided that he performs optimal measurements.  By varying the input state one can search for an optimal strategy, i.e., the initial state and measurement observables which minimise the probability of misidentification.

This approach was used for the purpose of detecting lossy channels \cite{invernizzi}, improving target detection \cite{Tan2008} and boosting the readout of digital memories \cite{Pirandola2011}. Here we show, using the Unruh theory as a specific example, that quantum channel discrimination can also be applied to test physical theories. We will show that the Unruh theory and the null theory can be thought of as two different quantum channels. Therefore deciding which theory is correct maps to the problem of discriminating which of these two channels operate when changing from an inertial to an accelerated frame. Our objective is to determine which initial state should be sent down this unidentified channel (which state should be prepared in the inertial frame) and which observables should be measured at the channel output (which observables should be measured in the accelerated frame) such that the actual channel (the correct theory) can most clearly be revealed.

Ordinarily one attempts to verify the Unruh effect by measuring particles in the vacuum from an accelerated frame. The detection of any number of particles would be evidence in favour of the Unruh theory. However, such a test is not perfect. Even excluding the possibility of dark counts, a thermal state is not orthogonal to the vacuum state.  Therefore in such experiments there is always some probability of making an error, for example by (incorrectly) identifying the vacuum when in actuality the state was thermal. The question then is, do other strategies exist which reduce these identification errors?

In this article we answer this question in the affirmative and report on feasible strategies that can be used to discriminate the Unruh theory that outperform this simple vacuum particle counting approach. These results are expected to be useful in tests of the Unruh theory in analogue experiments that are due to come online in the near future.   This provides a proof of principle that these tools can also be used, for example, in testing Hawking radiation in analogue systems \cite{Analogue} and other such tests of quantum field theory in curved spacetimes.

The outline of the paper is as follows: we first give some background on the Unruh effect in section \ref{sec:background} and present some mathematical definitions that will be of relevance to later sections in the paper. We then discuss an alternative theory to the Unruh theory in section \ref{sec:alternative} which does not predict the appearance of particles in the accelerated frame. We show in section \ref{E0E1channels} that the effect of these theories is to transform states in the inertial frame into states in the accelerated frame, and that these transformations are naturally described by quantum channels. After that we introduce in section \ref{sec:qcd} the subject of quantum channel discrimination and then use it in section \ref{displaced} to determine the optimal experimental setting to discriminate the theories with an initial coherent state.  We then investigate in section \ref{sec:optGaussian} other states starting with general Gaussian states that take the form of an Unruh mode, and then Fock states in section \ref{sec:Fock} for general initial states that are measured in a quasimonochromatic frequency band.  Finally we compare strategies for initially quasimonochromatic modes in section \ref{sec:realistic} before finishing with some concluding discussions.

There are five appendices. In Appendix \ref{app:orthobases} we provide the construction of non-standard orthonormal bases for Rindler and Minkowski frames. In Appendix \ref{app:FormalDerChannels} we provide further details on the channels associated with the two hypotheses that we consider in moving into the accelerated frame. In Appendix \ref{sec:XYderive} we derive the Gaussian channel matrices for the Unruh channel and in Appendix \ref{app:Fock} we derive the channels for Fock states before giving some further information on the numerical implementation of our realistic mode numerical calculations in Appendix \ref{app:numeric}.

\section{Background} \label{sec:background}
Our analysis is presented for a real massless scalar field in 1+1 dimensions ($\hbar=c=k_B=1$), but it can be generalised to any specific experimental setup. The Klein-Gordon equation is:
\bea
D^{\mu}\partial_{\mu}\phi=0,
\eea
where $D^{\mu}$ is the covariant derivative, and there exists a natural indefinite product on the space of solutions to this equation, called the Klein-Gordon scalar product, given by \cite{convention}:
\bea\label{scalarproduct}
(\phi_1,\phi_2)\equiv i \int \phi_1^{\star} \stackrel{\leftrightarrow}{\partial^{\mu}}\phi_2d\Sigma_{\mu}.
\eea

Throughout this paper we will call any solution of the Klein-Gordon equation a ``mode.'' In particle physics the terminology is usually reserved for the energy eigenmodes of the system.  However, in this paper the distinction of whether the solution is an energy eigenmode or a wavepacket of such modes is intentionally left ambiguous. This is in part because there are two Killing vectors which are time-like in the left and right wedge restriction of 1+1 Minkowski spacetime: the energy operator $\hat{E}=i\frac{\partial}{\partial t}$ and the boost operator $\hat{K}=i(x\frac{\partial }{\partial t}+t\frac{\partial}{\partial x})$. Hamiltonians can be defined on spacetime regions when there exists a time-like Killing vector (see discussion on page 15 of \cite{Takagi}). For each time-like Killing vector on a spacetime region there is a corresponding Hamiltonian.  There are therefore two different definitions of Hamiltonian in the left and right wedge restriction of 1+1 Minkowski spacetime. Since an eigenmode with respect to one operator may be a wavepacket of eigenmodes with respect to the other the special terminology of calling an energy eigenfunction a ``mode'' is not very illuminating. Another reason for our choice of terminology is that it is already prevalent in quantum optics to call a wavepacket a mode, and it will be familiar to those readers.

In Minkowski coordinates, the eigenfunctions of the ordinary energy operator, $\hat{E}$, are plane waves and are given by:
 \bea \label{MinkPlaneWaves}
u_k(x,t)\equiv \frac{1}{\sqrt{4\pi |k|}}e^{i(k x- |k| t)}.
\eea
In quantum optics experiments with resting detectors, it is common practice to analyse these frequencies using filters.  In this paper, we will consider an equivalent experiment with an accelerating particle detector. We suppose that an inertial source shines radiation onto an accelerating detector that makes measurements using filters in the accelerated frame.

We suppose that the detector follows a $\xi=0$ trajectory in Rindler coordinates $(\tau,\xi)$ which are related to time $t$ and position $x$ by:
\bea
 t&=&a^{-1}e^{a\xi}\sinh{a\tau},\\
 x&=&a^{-1}e^{a\xi}\cosh{a\tau},
 \eea
 where $a$ is the detector's proper acceleration and $\tau$ is the proper time along the trajectory. In these coordinates the boost operator becomes, $\hat{K}=\frac{i}{a}\frac{\partial}{\partial \tau}$. The actual Hamiltonian associated with this time-like Killing vector is $i\frac{\partial}{\partial \tau}$, or  $a \hat{K}$. However, henceforth we will simply refer to $\hat{K}$ itself as the energy operator, and the interpretation should be clear from the context.

 Canonical quantisation of the scalar field on the Rindler line-element, $ds^2=e^{2a\xi}(d\tau^2-d\xi^2)$, has been discussed by Fulling \cite{Fulling1972} leading to a different vacuum to the Minkowski vacuum called the Rindler vacuum, $|0\rangle_\text{I}$. By fitting two Rindler coordinate patches to cover the left and right wedges, Unruh has found \cite{Unruh1976} (see also \cite{Bruschi2010}) a relation  between the Minkowski vacuum state,  $|0\rangle_\text{M}$, and the product of the left and right wedge Rindler vacua, $|0\rangle_\text{R}\equiv |0\rangle_\text{I}\otimes |0\rangle_\text{II}$,  given symbolically \cite{UnitaryInequiv} by:
\bea\label{RindlerToMink}
 |0\rangle_\text{M}\propto\hat{S}_\text{I,II}|0\rangle_\text{R},
\eea
where the squeezing operator, $\hat{S}_\text{I,II}$,  is characterised by the squeezing parameter, $r_k={\rm arctanh}(e^{-\pi |k|/a})$, and fulfils the following relations:
\bea
\label{commut1}
\hat{S}_\text{I,II}\hat{b}_{\text{I}k}\hat{S}^{\dagger}_\text{I,II}&=&\cosh{r_k}\hat{b}_{\text{I}k}-\sinh{r_k}\hat{b}^\dagger_{\text{II}k},\\
\hat{S}_\text{I,II}\hat{b}_{\text{II}k}\hat{S}^{\dagger}_\text{I,II}&=&-\sinh{r_k}\hat{b}^{\dagger}_{\text{I}k}+\cosh{r_k}\hat{b}_{\text{II}k},\label{commut2}
\eea
where $\hat{b}_{\text{I}k}$ and $\hat{b}_{\text{II}k}$ are the annihilation operators associated with the Rindler modes:
\bea \label{rindlerI}
w_{\text{I}k}(\xi,\tau)=\frac{1}{\sqrt{4 \pi |k|}}\exp{i(k\xi-|k|\tau)},
\eea
and
\bea \label{rindlerII}
w_{\text{II}k}(\xi',\tau')=\frac{1}{\sqrt{4 \pi |k|}}\exp{i(-k\xi'-|k|\tau')},
\eea
respectively. In the left wedge we have used the coordinate patch:
\bea
t&=&a^{-1}e^{a\xi'}\sinh{a\tau'},\\
x&=&-a^{-1}e^{a\xi'}\cosh{a\tau'},
\eea
and in this wedge the boost operator can be written $\hat{K}=-\frac{i}{a}\frac{\partial}{\partial \tau'}$.

When a trace is performed over the Rindler modes in the left wedge of the Minkowski vacuum state a thermal state is obtained at a temperature proportional to the acceleration.

\section{An alternative hypothesis of non-inertial motion}\label{sec:alternative}
Any physically realised detector used by the accelerated observer to measure the radiation will have a limited bandwidth over which signals can be detected. To simplify our discussion we assume that frequencies can be selected by the detector by placing in front a linear filter such as a Fabry-Perot interferometer, or by ``homodyning" with a specific local oscillator mode \cite{Downes2013}. Since the detector is accelerated, the frequencies selected are defined with respect to the proper time of the accelerated observer, $\tau$, i.e., they are eigenfunctions of the boost operator.  Somewhat surprisingly this criteria alone does not uniquely define the physics in the accelerated frame.

Linear superpositions of the Rindler mode functions (\ref{rindlerI}), (\ref{rindlerII}) and their complex conjugates, can be taken to find  other solutions to the $\hat{K}$ eigenvalue equation. Of particular importance are the solutions known as Unruh modes. There are two types of Unruh modes, called Right-Unruh modes and Left-Unruh modes, and unlike the Rindler modes, which are localised in the left and right wedges, Unruh modes are distributed throughout all of space. Note that the Right (Left) prefix is not to be confused with right-moving (left-moving) waves, rather the prefix is supposed to indicate that the mode is mostly distributed within the right (left) wedge; for each type of Unruh mode, left-moving and right-moving solutions exist which are distinguished by the sign of $k$. Explicitly, the positive norm Unruh modes are:
\bea
u_{\text{R}k}(x,t)&=&\frac{1}{\sqrt{4 \pi |k|}}\frac{1}{\sqrt{\epsilon(1-e^{-2 \pi k/a})}}(a(\epsilon x-t))^{i k /a},\\
u_{\text{L}k}(x,t)&=&\frac{1}{\sqrt{4 \pi |k|}}\frac{1}{\sqrt{\epsilon(e^{2 \pi k/a}-1)}}(a(\epsilon x-t))^{-i k /a},
\eea
where $\epsilon\equiv \text{sign}(k)$, $\log{(-1)}=i\pi$ and the branch cut is taken in the lower-half complex plane, below the negative real axis. Furthermore, the negative norm Unruh modes are:
\bea
u_{\text{R}k}(x,t)^{\star}&=&\frac{1}{\sqrt{4 \pi |k|}}\frac{1}{\sqrt{\epsilon(e^{2 \pi k/a}-1)}}(-a(\epsilon x-t))^{-i k /a},\\
u_{\text{L}k}(x,t)^{\star}&=&\frac{1}{\sqrt{4 \pi |k|}}\frac{1}{\sqrt{\epsilon(1-e^{-2 \pi k/a})}}(-a(\epsilon x-t))^{i k /a}.
\eea
It should be noted that $|k|$ is the eigenvalue of the Hamiltonian associated with $\hat{K}$.

The Unruh modes are related to the Rindler modes by the simple equations:
\bea \label{UnruhRindler1}
u_{\text{R}k}& =&\cosh r_k w_{\text{I}k}+\sinh r_k w_{\text{II}k}^{\star},\\
u_{\text{L}k}^{\star}& =&\sinh r_k w_{\text{I}k}+\cosh r_k w_{\text{II}k}^{\star} .\label{UnruhRindler2}
\eea
It should be clear that Unruh modes and Rindler modes coincide up to different normalisation factors in the left and right wedges respectively. By associating operators $\hat{A}_{\text{R}k}$ and $\hat{A}_{\text{L}k}$ with the Right-Unruh and Left-Unruh modes respectively, we obtain the operator relations:
\bea\label{UnruhRindlerOps1}
\hat{A}_{\text{R}k}& =&\cosh r_k \hat{b}_{\text{I}k}-\sinh r_k \hat{b}_{\text{II}k}^{\dagger},\\
\hat{A}_{\text{L}k}& =&-\sinh r_k \hat{b}_{\text{I}k}^{\dagger}+\cosh r_k \hat{b}_{\text{II}k}.\label{UnruhRindlerOps2}
\eea

One interesting feature of Unruh modes is that they have a definite frequency property with respect to both $\hat{K}$ and $\hat{E}$. For example, $u_{\text{R}k}$ is a positive frequency eigenfunction of $\hat{K}$ but it can also be decomposed only in terms of positive frequency eigenfunctions of $\hat{E}$. On the other hand for example, the $w_{\text{I}k}$ and $w_{\text{II}k}^{\star}$ Rindler modes are positive frequency with respect to $\hat{K}$, but mixed with respect to $\hat{E}$,  that is, they are superpositions of both positive and negative frequency eigenfunctions of $\hat{E}$. A summary of the frequency properties of the special modes considered in this paper is shown in table \ref{tab:modeprops}.

\begin{table}[t]
\begin{tabular}{c|ccc}
Mode                              & $\hat{E}$ & $\hat{K}$ & Norm \\
\hline\hline
$u_k$                             & $+$           & ``mixed''   & $+$\\
$u_k^{\star}$                  & $-$           & ``mixed''     & $-$\\
$w_{\text{I}k}$                & ``mixed''  & $+$              & $+$\\
$w_{\text{I}k}^{\star}$     &``mixed''   & $-$               & $-$\\
$w_{\text{II}k}$                & ``mixed'' & $-$              & $+$\\
$w_{\text{II}k}^{\star}$     &``mixed''  & $+$              & $-$\\
$u_{\text{R}k}$                & $+$        & $+$              & $+$\\
$u_{\text{R}k}^{\star}$     &$-$          & $-$               & $-$\\
$u_{\text{L}k}$                & $+$        & $-$              & $+$\\
$u_{\text{L}k}^{\star}$     &$-$         & $+$              & $-$
\end{tabular}
\caption{\label{tab:modeprops} Frequency and norm \cite{symbolconvention} properties of the Minkowski, $u$, Rindler $\{w_\text{I}$, $w_\text{II}\}$ and Unruh $\{u_\text{R}$, $u_\text{L}\}$ modes (defined in the text) and their complex conjugates. A mode is said to have a positive (negative) frequency property with respect to the energy operator $\hat{E}$ or $\hat{K}$ if it can be expressed as a superposition of only positive (negative) frequency eigenfunctions of that operator. If it can not it is labelled as ``mixed.'' }
\end{table}

The Hilbert space quantisation critically depends on the frequency properties of the modes. However, we have seen that one cannot uniquely define the Hilbert space with respect to the positive frequencies of the boost operator:  the degeneracy of the space of positive frequency solutions with a definite eigenvalue (and definite parity) is two dimensional.  If we let $\alpha$ and $\beta$ be two complex numbers such that $|\alpha|^2+|\beta|^2=1$, then the superpositions of positive-norm Right Unruh modes and negative-norm Left Unruh modes:
\bea
f_{k}=\alpha u_{\text{R}k}+\beta u_{\text{L}k}^{\star},
\eea
are also positive frequency with respect to the boost operator. Clearly there are an infinity of possible solutions satisfying the positive frequency criteria, therefore further assumptions are necessary to lead to a unique physical outcome.

The standard choice is obtained by setting $\alpha=\cosh r_k$ and $\beta=-\sinh r_k$ leading to a right-wedge Rindler mode. Such a choice arises naturally when working in Rindler coordinates and leads to the well-known results of Unruh \cite{Unruh1976}.  On the other hand, Unruh modes also play a special role, they are the unique set of eigenfunctions of $\hat{K}$ that have a definite frequency property with respect to $\hat{E}$.  This is noteable because ordinary quantum field theory in the inertial frame distinguishes those modes that are positive frequency with respect to $\hat{E}$, and we are interested in verifying that this distinguished role is not respected in the uniformly accelerated frame, as suggested by the standard Unruh theory.

An accelerated particle detector may respond to the Hilbert space defined by those frequencies which are positive frequency with respect to $\hat{E}$, namely to the $u_{\text{R}k}$ and $u_{\text{L}k}$ solutions, or to those which are not positive frequency with respect to $\hat{E}$, of which there are an infinity of possibilities.  Standard theory dictates that the correct choice of solutions in the latter case are the right-wedge Rindler modes $w_{\text{I}k}$ (note we are only considering a detector in the right wedge).

We are therefore interested in experimentally determining which of these two situations, if any, occurs in practice. It is of course possible to test the other alternatives, however for the reasons we have outlined above we find these two cases to be the most compelling.  Under the assumption that one of these two alternatives is correct it is possible to frame the problem of determining which of the two is correct in terms of a binary hypothesis test.

We label the alternative theory as H0 for null theory, and the standard Unruh theory as H1. In quantum hypothesis testing the choice of H0 and H1 is symmetrical. In particular, we are not implying that H0 should be accepted as correct until proven otherwise as might be the case in standard hypothesis testing.

The hypotheses can be briefly surmised as follows:
\begin{itemize}
\item \textbf{H0 (The null theory):} Under the H0 hypothesis the detection modes are hypothesised to have the positive frequency property with respect to $\hat{E}$. Therefore, they are wavepackets of $u_{\text{R}k}$ and $u_{\text{L}k}$ \cite{uLjustify}. We label this wavepacket by $\psi_0$, and note that the operator associated with it will annihilate the Minkowski vacuum state.

\item \textbf{H1 (The Unruh theory):} Under the H1 hypothesis the detection modes are hypothesised to be wavepackets of right-wedge Rindler modes, $w_{\text{I}k}$ (i.e., with support on $x>|t|$ only). These modes do not have the positive frequency property with respect to $\hat{E}$. Therefore, the operators associated with these wavepackets do not annihilate the Minkowski vacuum (rather they annihilate the Rindler vacuum). Particles in these modes, first discussed by Fulling \cite{Fulling1972}, are called Rindler-Fulling particles. In the Unruh theory we will consider a wavepacket of right-wedge Rindler modes as the detection mode, and label this wavepacket by $\psi_1$.
\end{itemize}

One may wish to imagine the detector as a harmonic oscillator whose Hilbert space is taken to be a subspace of (and shares its ground state with) a Fock space describing the quantum field. Hypothesis H0 is that this Fock space has one-particle space comprising positive frequency (w.r.t. both $\hat{E}$ and $\hat{K}$) Unruh modes, among which is the excitation mode of the oscillator; hypothesis H1 is that the Fock space has one-particle space comprising positive $\hat{K}$-frequency Rindler modes, among which is the excitation mode of the oscillator.

Since the vacuum state of the detector in the H0 hypothesis shares the same vacuum state as the Minkowski vacuum state, the H0 hypothesis leads to the alternative conclusion that an accelerated observer would not detect particles in the Minkowski vacuum in agreement with an Unruh-effect skeptic \cite{Narozhny, Comment}.

\section{Two quantum channels of non-inertial motion}
\label{E0E1channels}
In the last section we showed that there is a meaningful way of defining a null theory which contains the prediction that the vacuum state measured in the accelerated frame is void of particles. However, we can do more than simply investigate the vacuum state. We can also ask what predictions the H0 hypothesis makes for other initial states of the field. In the H0 case, the only reason that the measured state of the field is not given trivially by the initial state itself, is because of our measurement assumptions: we assume that there is a finite bandwidth of frequencies which the detector can measure. In the accelerated frame this is a bandwidth in $\hat{K}$ space.  Therefore, the map which takes the initial state of the field to the subspace measurable by the detector is obtained by a change from the standard Minkowski basis into the Unruh mode basis, followed by a trace over all Unruh modes in the inaccessible part of the Hilbert space, i.e., those frequencies that are out of range.

The situation is not so different in the H1 hypothesis. However, there is a new feature arising because of the different vacua, known as amplification. Amplification occurs when a process creates particles. In the H1 hypothesis, the measured frequencies are assumed to be right-wedge Rindler modes. So following in the same fashion as before, we rewrite the initial state this time into the Rindler basis, and then trace out all modes that are out of range. However, included in this set of out-of-range modes are the left-wedge Rindler modes. Because of the nature of the Minkowski vacuum state, these modes will in general be highly entangled with the modes in the right-wedge. In particular, negative energy modes are paired with positive energy modes, so when the left wedge is traced out there is the appearance of particle creation in the right wedge.

We call the maps which take the input state of the field to the state measured by the detector (or if one prefers, to the state of the detector itself) $\mathcal{E}_0$ and $\mathcal{E}_1$, which are labeled after the H0 and H1 hypotheses respectively. We will hereafter refer to these maps as channels \cite{channeldef}. We have argued that these channels take the form:

\bea\label{E0general}
\mathcal{E}_0(\rho)&=&\text{Tr}_{\perp \psi_0}[U_0\rho U_0^{\dagger}],\\
\mathcal{E}_1(\rho)&=&\text{Tr}_{\perp \psi_1}[U_1\rho U_1^{\dagger}],\label{E1general}
\eea
where $U_0$ ($U_1$) is a transformation operator from the Minkowski basis into the Unruh (Rindler) basis, and $\text{Tr}_{\perp \psi}$ means trace out all modes orthogonal to the $\psi$ subspace.  We provide more details on these relations in appendix \ref{app:FormalDerChannels}.

To gain a better insight into the nature of these channels it is useful to consider the special case when $\psi_0=u_{\text{R}k}$ is a Right-Unruh mode and $\psi_1=w_{\text{I}k}$ is a right-wedge Rindler mode. In this case, the H0 state, $\mathcal{E}_0(\rho)$, is simply a state defined on the single mode subspace of $u_{\text{R}k}$. By virtue of equation  (\ref{UnruhRindler1}) this is very nearly the same state as the H1 state, $\mathcal{E}_1(\rho)$. The difference is a subsequent unitary squeezing operation on $\mathcal{E}_0(\rho)$ that changes the state into the Rindler mode basis, followed by a trace over the $w_{\text{II}k}$ mode subspace. Since the explicit operator for the unitary squeezing operation on the Unruh subspace is:
\bea
S=e^{r_{k} (\hat{b}_{\text{I}k}^{\dagger}\hat{b}_{\text{II}k}^{\dagger}-\hat{b}_{\text{I}k}\hat{b}_{\text{II}k}) },
\eea
we can write:
\bea
\mathcal{E}_1(\rho)=\text{Tr}_{\text{II}} [S\mathcal{E}_0(\rho)S^{\dagger}],
\eea
where the trace is performed over the subspace defined by $w_{\text{II}k}$.

One way of defining a Bosonic amplification channel \cite{Aspachs, Unruh2011, Alsing2004} is via the map $\mathcal{E}_\text{amp}(\rho)\equiv \text{Tr}_{\text{II}} [S\rho S^{\dagger}]$, (see, for example, the discussion in the first column of pg. 2 in \cite{Aspachs}). Therefore, the H1 channel, $\mathcal{E}_1$, can be decomposed into an $\mathcal{E}_0$ channel followed by a Bosonic amplification channel, i.e., $\mathcal{E}_1=\mathcal{E}_\text{amp}\circ\mathcal{E}_0$. We can see here that the test we are performing is really whether or not the amplification channel is operating. Indeed, it is the amplification channel which leads to the observation of particles in the accelerated frame and it is what we consider to be the most profound aspect of the theory -- the property that we most want to test.

In our discussion so far we have implicitly assumed that the initial state is simple in the standard Minkowski basis. By `simple' we mean the excitations above the vacuum of any prepared state have sharp $\hat{E}$-frequencies (these are known as a quasimonochromatic modes).  However, in principle the experimentalist is at liberty to tailor the mode prepared by the source to suit the experimental purpose. One might then wonder if there was a preferred mode shape in which the experimentalist could prepare the initial state such that the amplification would most clearly be revealed. Indeed, when the source mode itself is an Unruh mode the $\mathcal{E}_\text{0}$ map becomes trivial, $\mathcal{E}_\text{0}\rightarrow\mathcal{I}$. If it were possible to prepared an Unruh mode, then the Unruh effect could be tested by discriminating between an amplification channel and a trivial channel.  Despite the fact that it is not currently known how to produce such modes, much insight can be gained by first studying this simpler scenario and we will present some results for initial Unruh modes in the following sections. However, to make closer connection to settings that are likely to be experimentally feasible, we also consider source modes that are peaked in ordinary Minkowski frequencies by taking narrow spectrally-uniform wavepackets \cite{Hawking1975, Takagi} (see for example pg. 18 of \cite{Takagi}). In this case, the $\mathcal{E}_\text{0}$ channel is no longer trivial.

\section{Quantum channel discrimination}\label{sec:qcd}
We try several different input states, $\rho$, and assume that $N$ identical copies of each of them are available for collective measurement.  Quantum state discrimination is then performed on the two output states $\mathcal{E}_0(\rho)$ and $\mathcal{E}_1(\rho)$ corresponding to each of the hypotheses.

Quantum state discrimination can be implemented by measuring a two-outcome positive operator valued measurement with operators $E_{0}$ and $E_{1}$, satisfying $E_{0}+ E_{1}=\mathbb{I}$ and $E_{i}\geq0 ~\forall i$.

The outcomes of these measurements are assigned to different interpretations of the theories as follows. If the outcome $E_0$ is obtained one infers that the theory H0 is correct. On the other hand if the outcome $E_{1}$ occurs one infers that the H1 hypothesis is correct.

The probability of misidentification of a given strategy is given by the weighted sum of the probability of measuring $E_0$ when the H1 hypothesis is correct and the probability of measuring $E_1$ when the H0 hypothesis is correct, where the weights are given by the a priori probabilities for each of the hypothesises of being correct. Using the fact that the operators must sum to the identity one arrives at the total error probability of misidentification:
\bea \label{probmisid}
P_\text{err}=\frac{1}{2} (1-\text{Tr}\left[E_1\Lambda\right]),
\eea
 where
 \bea \label{helstrom}
 \Lambda\equiv\mathcal{E}_1(\rho)-\mathcal{E}_0(\rho),
 \eea
 is the Helstrom matrix. The a priori probabilities for each of the hypotheses have been assumed to be equal to one-half. Optimising over all positive operator valued measurements one obtains the Helstrom bound  \cite{QSD}:
\bea
P_\text{hel}=\frac{1}{2}-\frac{1}{4}|\Lambda|.
\eea
Note that the norm here refers to the operation of taking the sum of the absolute values of the eigenvalues.

Consider now the simplest example, when the initial state is the vacuum. Then it follows that the H0 state is also the vacuum, $\mathcal{E}_0(\rho)=|0\rangle\langle 0|$. On the other hand,  the H1 state is a thermal state (i.e., the Unruh thermal state): $\mathcal{E}_1(\rho)=\frac{1}{n+1}\sum_m\left(\frac{n}{n+1}\right)^m|m\rangle\langle m|$, where henceforth $n$ is defined as the mean particle number in the detection mode, $\psi_1$, when the initial state is the Minkowski vacuum \cite{Dragan2012}. This can be expressed in terms of the right-wedge Rindler mode (\ref{rindlerI}):
\bea\label{noise}
n=\int d|k| \frac{|(\psi_1,w_{\text{I}k})|^2}{e^{2\pi |k|/a}-1}.
\eea
Since the Helstrom matrix is diagonal and only the first eigenvalue is negative, we immediately deduce that the optimal measurement is $E_0=|0\rangle\langle 0|$ and  $E_1=\mathbb{I}-|0\rangle\langle 0|$, which is simply a test of the existence of particles. In this strategy the probability of misidentification is:
\bea \label{p0}
P_0=\frac{1}{2(n+1)}.
\eea
One should recognise this as the strategy to observe the Unruh effect described in the introduction. Our objective is to find alternative strategies that reduce this probability of error thereby enhancing discrimination of the theories.

It is not always possible to calculate the Helstrom bound exactly. A more readily computable upper bound is the Quantum Chernoff Bound (QCB), $P_{\text{QCB}}^{(N)}$, \cite{Audenaert2007}:
\bea\label{helQCBbound}
P_{\text{hel}}^{(N)}\leq P_{\text{QCB}}^{(N)}\equiv \tfrac{1}{2} \text{exp}{(-\kappa N)},
\eea
where $N$ is the number of independent copies of the state and
\bea
\kappa\equiv -\ln\left[\inf_{0\leq s\leq 1} \text{Tr}(\mathcal{E}_0^s(\rho)\mathcal{E}_1^{1-s}(\rho))\right],
\eea
is the quantum Chernoff information giving the exponent for which the probability of misidentification most quickly decreases with increasing $N$.   In practice multiple independent copies of the state would be used to discriminate the theories. In the limit that $N\rightarrow\infty$ the inequality in (\ref{helQCBbound}) becomes tight. Therefore, in the asymptotic limit finding the state which minimises the QCB is equivalent to optimising the strategy. Since minimisation of the QCB over the single copy state implies minimisation over the multiple copy state \cite{invernizzi} (constrained by mean energy per copy), we only need to perform the analysis of the single copy state. It should be clear that by \textit{optimal state} we mean the state which minimises the QCB and therefore provides the minimum error probability in the asymptotic limit of many copies. Furthermore, in this limit,
the QCB bound does not depend on the a priori probabilities of H0 and H1 \cite{Audenaert2008}, which can be then considered completely
arbitrary. For calculating the QCB we use the tools and conventions of \cite{Pirandola2008}. \newline

\section{Single mode displaced vacuum states}\label{displaced}
We first consider probing the channels with a coherent state
\bea
|\alpha\rangle\equiv e^{\alpha\hat{A}^{\dagger}-\alpha^{\star}\hat{A}}|0\rangle,
\eea
with mean particle number $n_0=|\alpha|^2$. Note that the coherent state is in a general mode $\phi$ with corresponding annihilation operator $\hat{A}=(\phi,\hat{\Phi})$, where $\hat{\Phi}$ is the canonically quantised field operator and $(\cdot,\cdot)$ is the Klein-Gordon scalar product (\ref{scalarproduct}).  Since the initial state is Gaussian and both  $\mathcal{E}_0$ and  $\mathcal{E}_1$ are Gaussian channels, the output states are fully described by their first and second statistical moments.

We have shown in appendix \ref{sec:XYderive} that $\mathcal{E}_1(\rho)$ is a displaced thermal state, with thermal number $n$, and displacement $\alpha_1\equiv (\psi_1,\phi) \alpha+ (\psi_1,\phi^{\star}) \alpha ^{\star}$. Similarly, $\mathcal{E}_0(\rho)$ is found to be a coherent state, with displacement $\alpha_0\equiv (\psi_0,\phi)\alpha $. Since $\mathcal{E}_0(\rho)$ is pure, the QCB reduces to the fidelity, $\mathcal{F}$, \cite{Spedalieri} and the probability of error is:
\begin{eqnarray}
P_\text{hel}^{\text{coh}}\leq\tfrac{1}{2}{\mathcal{F}\Big(\mathcal{E}_0(\rho),\mathcal{E}_1(\rho)\Big)}.
\end{eqnarray}
The fidelity is a measure of the ``closeness'' of two quantum states. The fidelity between two single-mode Gaussian states $\rho _{A}$ and $\rho _{B}$, with moments $(%
\mathbf{V}_{A},\mathbf{\bar{x}}_{A})$ and $(\mathbf{V}_{B},\mathbf{\bar{x}}%
_{B})$, is given by the formula  \cite{fidelity,Weedbrook2012}:%
\begin{equation}
\mathcal{F}(\rho _{A},\rho _{B})=\frac{2}{\sqrt{\Delta +\delta }-\sqrt{\delta }}\exp %
\left[ -\tfrac{1}{2}\mathbf{d}^{T}(\mathbf{V}_{A}+\mathbf{V}_{B})^{-1}%
\mathbf{d}\right] ,  \label{Fidelity_pairGaussian}
\end{equation}%
where%
\begin{equation}
\Delta \equiv\det (\mathbf{V}_{A}+\mathbf{V}_{B})~,~\delta \equiv(\det \mathbf{V}%
_{A}-1)(\det \mathbf{V}_{B}-1)~,  \label{DeltasFORfidelity}
\end{equation}%
and $\mathbf{d}\equiv\mathbf{\bar{x}}_{A}-\mathbf{\bar{x}}_{B}$.
In this notation, for the two states $\mathcal{E}_0(\rho)$, $\mathcal{E}_1(\rho)$ we find that: $\delta=0$, $\Delta=(2n+2)^2$, and $\mathbf{V}_{A}+\mathbf{V}_{B}=(2n+2)\mathbb{I}_2$ where $\mathbb{I}_2$ is the $2\times2$ identity matrix.

Therefore, the QCB can be written:
\begin{eqnarray}
\label{csQCB}
P_\text{hel}^{\text{coh}}&\leq& \frac{1}{2(n+1)}\text{exp}\left({-\tfrac{\mathbf{d}^T\mathbf{d}}{2(2n+2)}}\right)\\
&=&P_0\text{exp}\left({-\tfrac{|\alpha_1-\alpha_0|^2}{(1+n)}}\right),
\end{eqnarray}
where on the last line we used equation (\ref{p0}) and:
\bea
\mathbf{d}=2\left(\begin{array}{c} \text{Re} [(\alpha_1-\alpha_0)]\\
\text{Im} [(\alpha_1-\alpha_0)]
\end{array}\right).
\eea

Therefore, coherent probes enhance the discrimination of the Unruh effect in a fashion which scales exponentially with the energy.

The strategy which achieves this probability of error corresponds to the measurement observables $E_0=|\alpha_0\rangle \langle \alpha_0|$ and $E_1=\mathbb{I} -|\alpha_0\rangle \langle \alpha_0|$. To see this, we calculate the probability of error of this strategy from equation (\ref{probmisid}):
\bea
P_\text{ken}&=& \frac{1}{2}\left(\text{Tr}[E_1\mathcal{E}_0(\rho)]+\text{Tr}[E_0\mathcal{E}_1(\rho)]\right)\\
&=& \frac{1}{2} \text{Tr}[\mathcal{E}_0(\rho)\mathcal{E}_1(\rho)].
\eea
Note that when either of $\mathcal{E}_0(\rho)$ or $\mathcal{E}_1(\rho)$ are pure $\text{Tr}[\mathcal{E}_0(\rho)\mathcal{E}_1(\rho)]$ is equal to the fidelity. Therefore this probability of error is equal to the QCB. In practice, this strategy is performable with a Kennedy receiver, see pg. 15 of \cite{Weedbrook2012}: first one displaces the state by $D(-\alpha_0)$ and then performs a measurement of whether or not there are particles, i.e., the $E_0=|0\rangle\langle0|$, and  $E_1=\mathbb{I}-|0\rangle\langle 0|$ (This last step can be easily done with a simple particle counting detector).

We have shown that the Kennedy receiver achieves the QCB. While in general this is not the optimal measurement strategy,  we know that in the limit of many repetitions of the experiment the QCB becomes tight.

\section{Optimized discrimination with Gaussian states} \label{sec:optGaussian}
One naturally wonders how close the strategy presented in the previous section comes to being optimal. To find the optimal state, one must perform an exhaustive search over all initial states at fixed energy. Since this is not practical in full generality, we first focus on special classes of Gaussian states. General non-unitary Gaussian transformations can be written as a transformation of the first and second moments, $\overline{\mathbf{x}}'=X\overline{\mathbf{x}}$, and $\overline{\mathbf{V}}'=X\overline{\mathbf{V}} X^{T} +Y$ respectively \cite{Serafini2005, Aspachs,Weedbrook2012}. For single-mode Gaussian states the Unruh channel is completely described by:
\bea
X_1&=&\left(\begin{array}{cc}
\text{Re}[(\psi_1,\phi)+(\psi_1,\phi^{\star})]&-\text{Im}[(\psi_1,\phi)+(\psi_1,\phi^{\star})] \\
\text{Im}[(\psi_1,\phi)-(\psi_1,\phi^{\star})] & \text{Re}[(\psi_1,\phi)-(\psi_1,\phi^{\star})]
\end{array}\right),\nonumber\\
Y_1&=& (2n+1) \mathbb{I}_2-X_1 X_1^{T}.\label{H0GausTransfs}
\eea
These transformation matrices have been derived in appendix \ref{sec:XYderive}. Similarly, the H0 transformations $\{X_0,Y_0\}$ are obtained by replacing $\psi_1\rightarrow \psi_0$ and $n\rightarrow 0$ in (\ref{H0GausTransfs}) \cite{beta0}.

In the remaining part of this section our results will focus on the special case when Alice prepares the state in an Unruh mode, more general results will be presented in section \ref{sec:realistic}.  When the initial state is an Unruh mode and the Rindler detection mode is tuned to the same $\hat{K}$-frequency, we obtain $(\psi_1,\phi)=\sqrt{n+1}$,   $(\psi_1,\phi^{\star})=0$ and $(\psi_0,\phi^{\star})=1$.  Equation (\ref{H0GausTransfs}) then reduces to $X_1=\sqrt{1+n}  \mathbb{I}_2$ and $Y_1=n  \mathbb{I}_2$ as in \cite{Aspachs} (see halfway down the second column on pg. 2).

\begin{figure}
\centering
\includegraphics[width=0.4\textwidth]{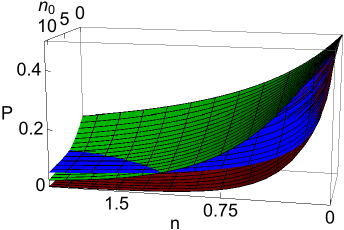}
\includegraphics[width=0.4\textwidth]{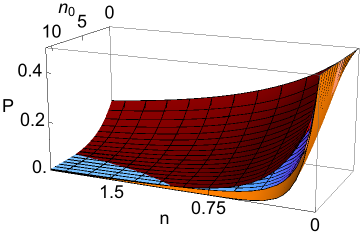}
\caption{\label{fig:Unruhplots} (Color Online) (Top) Comparison of the QCB error probability, $P$, for the single mode Gaussian states: coherent (green), squeezed (blue) and optimal single-mode gaussian state (red) in the Unruh mode scenario. (Bottom) Comparison of the optimised single-mode squeezed-displaced (red), two-mode optimised squeezed displaced (light blue) and Fock state (Orange). \vspace{-.5cm} }
\end{figure}

For the coherent state of the previous section the exponential factor becomes: $|\alpha_1-\alpha_0|^2=n_0(\sqrt{n+1}-1)^2$.  For a Single-Mode Squeezed vacuum state (SMS) with $n_0=\sinh^2{s_0}$, we find:
\bea
P_\text{hel}^{\text{sqz}}\leq
\frac{1}{2 \sqrt{1+n (2+n) (1+n_0)}},
\eea
which is proportional to the inverse square root of $n_0$ as $n_0\rightarrow\infty$.  Therefore, at large energies the coherent state beats the SMS. Nevertheless, at low energies the SMS provides an enhanced sensitivity particularly in the low temperature (i.e., low $n$) regime, see Fig.~\ref{fig:Unruhplots} (Top).

The Optimal Single Mode Gaussian state (OSMG) is found by considering a displaced squeezed thermal state of fixed energy, $n_0= \sinh^2{s_0}+m_0\cosh{2 s_0} + |\alpha|^2$, where $m_0$ is the thermal number of the initial state, and the state is displaced and squeezed in the $p$-quadrature direction \cite{QCBdirectionopt}. The energy budget for the squeezing energy, thermal energy and displacement energy is given by the ratios $\kappa_1=\sinh^2{s_0}/n_0$, $\kappa_2=m_0\cosh{2 s_0} /n_0$ and $\kappa_3=|\alpha|^2/n_0$ respectively, where $\kappa_1+\kappa_2+\kappa_3=1$, $\kappa_i\geq0$. We find the optimal value $\kappa_2=0$ for all values of the parameter space considered. Therefore, pure states are better probes of the $\mathcal{E}_1$ channel than mixed (thermal) states. The QCB error probabilities for the OSMG, coherent state and SMS are shown in Fig.~\ref{fig:Unruhplots} (Top).

To investigate the usefulness of entanglement, we also considered a two-mode one-party accelerated strategy, whereby an ancillary mode is entangled with the first but is measured in the inertial frame rather than the accelerated frame. The transformation matrices in this case are: $X\rightarrow \mathbb{I}_2\oplus X$, $Y\rightarrow \mathbb{I}_2\oplus Y$. The quantum correlations of a Two-Mode Squeezed vacuum state (TMS) under the one-party accelerated motion setting have previously been investigated in \cite{Doukas2012, Dragan2012, AdessoTMS}. Here we consider an initial two-mode squeezed state that is also displaced in the $x$-quadrature \cite{QCBdirectionopt}. The QCB optimised over displacement and squeezing is shown in Fig.~\ref{fig:Unruhplots} (Bottom).

In the low energy regime entanglement can be a useful resource in the discrimination. In particular, the two-mode Gaussian state that we have considered (optimised over displacement and squeezing) can beat the OSMG. However, the OSMG is still better for sufficiently large $n_0$.

It is important to mention that the measurement which obtains the QCB in the  two-mode one-party accelerated strategy would in general be non-local across both parties. While in practice this would be very difficult to achieve (since Alice and Bob are in different frames), our results set a lower bound on the error for any local measurement in this setup.

\section{Fock states}\label{sec:Fock}
Finally we consider the effects of non-Gaussianity by probing the channel with an $n_0$ particle Fock state, $\rho=|n_0\rangle\langle n_0|$, we find:
\begin{align}
\mathcal{E}_0(\rho)\!=&\!\!\sum_{k=0}^{n_0}
\binom{n_0}{ k}
\big|1-|(\psi_0,\phi)|^2\big|^{n_0-k} |(\psi_0,\phi)|^{2 k} |k\rangle\langle k |,\label{eqn:Fock1}\\
\mathcal{E}_1(\rho)\!=&\!\!\sum_{k=0}^{n_0}\sum_{i=0}^{\infty}
\binom{n_0}{ k}
\big|1-|(\psi_0,\phi)|^2\big|^{{n_0}-k} |(\psi_0,\phi)|^{2 k} C_{k,i}(n) |k\!\!+\!\!i\rangle\langle k\!\!+\!\!i |, \label{eqn:Fock2}
\end{align}
where $C_{k,i}(n)=\binom{k+i}{ k} (1+n)^{i-k-1}n^i $.  Note that the equations are only valid when $\psi_0$ (and hence $\psi_1$) is a quasimonochromatic mode in $\hat{K}$ space. These states generalise the ones found for Unruh modes in \cite{Aspachs} to general initial modes, $\phi$. The derivation of these states can be found in appendix \ref{app:Fock}.

When the initial mode is an Unruh mode the $\mathcal{E}_0$ channel is trivial (this also follows from (\ref{eqn:Fock1}) with $\psi_0=\phi$). Furthermore, the $\mathcal{E}_1$ channel simplifies to:
\bea
\mathcal{E}_1(\rho)=\sum_{i=0}^{\infty}C_{n_0,i}(n)| n_0+i \rangle \langle n_0+i|.
\eea
Therefore, the Helstrom matrix (\ref{helstrom}) can be written:
\bea
\Lambda
&=&(C_{n_0,0}(n)-1)|n_0\rangle\langle n_0| +\sum_{i=1}^{\infty}C_{n_0,i}(n)| n_0+i \rangle \langle n_0+i|.
\eea

Since this matrix is diagonal and $(C_{n_0,0}(n)-1)<0$ and  $C_{n_0,i}(n)>0$, it follows that the projector onto the positive eigenvalue subspace, and hence the optimal choice of $E_1$, is given by:
\bea
E_1=\mathbb{I}-|n_0\rangle\langle n_0|.
\eea
Consequently, $E_0=|n_0\rangle\langle n_0|$. Naturally, the optimal measurement corresponds to counting the number of particles and deciding H0 if the number of particles is exactly $n_0$ \cite{footFock} and H1 otherwise.

When the source mode is an Unruh mode we can also use (\ref{probmisid}) to find the Helstrom bound on the probability of error:
\bea
P_\text{n$_0$}=\frac{1}{2(n+1)^{n_0+1}}.
\eea

When the source mode is an Unruh mode, the Fock state strategy outperforms the OSMG and two-mode Gaussian strategies, see Fig.~\ref{fig:Unruhplots} (Bottom). Thus, if it were possible to prepare a Fock state in an Unruh mode, it would be the best strategy to experimentally discriminate the theories. Indeed, we see that increasing the initial number of particles gives an exponential improvement over the vacuum strategy.

However, this strategy is not optimal for general initial modes. To investigate the performance of Fock states in the general initial mode case, we will calculate the QCB using equation (\ref{helQCBbound}). This will require taking powers of $s$ and $1-s$ of the density matrices. Fortunately, equations (\ref{eqn:Fock1}-\ref{eqn:Fock2}) are already diagonal in the number basis. However, the minimisation over $s$, will need to be done numerically. In order to numerically handle the infinite sum, it is useful to rewrite equation (\ref{eqn:Fock2}) as follows:
\bea
\mathcal{E}_1(\rho)&=&\sum_{n'=0}^{\infty}\sum_{k=0}^{n_0}\sum_{i=0}^{\infty}
\binom{n_0}{ k}
\big|1-|(\psi_0,\phi)|^2\big|^{{n_0}-k} |(\psi_0,\phi)|^{2 k} \\
 \nonumber && \times \ C_{k,i}(n) |k+i\rangle\langle k+i | n'\rangle\langle n'|, \\
&=&\sum_{n'=0}^{\infty}\sum_{k=0}^{n'}
\binom{n_0}{ k}
\big|1-|(\psi_0,\phi)|^2\big|^{{n_0}-k} |(\psi_0,\phi)|^{2 k} \\ \nonumber && \times \ C_{k,n'-k}(n) |n'\rangle\langle n' |,
\eea
where on the first line we inserted a complete set of states, and on the last line we set $i=n'-k$ and made use of the fact that $i\geq0$ which implies $k\leq n'$. We can then take finite partial sums in $n'$ until the partial sums converge to required accuracy.

\section{A realistic example}\label{sec:realistic}
At low energies and Unruh temperatures Fock states and squeezed states clearly beat the coherent states in the Unruh mode setup. Since these states are readily produced in ordinary Minkowski frequency modes, could they be used to reveal the Unruh effect at low temperatures and low source energy?

Consider irradiating the detector with a quasi-monochromatic mode with a flat spectrum, a central frequency $\omega=a/10$ and a spectral width $\delta\omega=\omega/10$. Suppose that the detector response is also flat, operates at the Rindler frequency $\omega_\text{R}=a/10$ and has a spectral range $\delta\omega_\text{R}=\omega_\text{R}/10$. We calculate:
\bea
(\psi_0,\phi)\approx(\psi_1,\phi^{\star})&=&0.002+0.013 i,\\
(\psi_1,\phi)&=&0.003+0.017 i,
\eea
 and an expected vacuum particle number $n=1.07$. The details of these numerical computations can be found in appendix \ref{app:numeric}.

\begin{figure}
\centering
\includegraphics[width=0.41\textwidth]{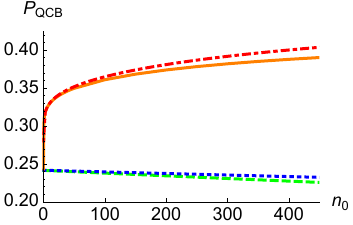}
\caption{\label{fig:realistic} (Color Online) QCB probabilities in the realistic mode example. Coherent state (green dashed), TMS vacuum state (blue dotted), SMS vacuum state (red dot-dashed) and Fock state (orange solid). }
\end{figure}
A comparison of the QCB for coherent, SMS, TMS, and Fock states is shown in Fig.~\ref{fig:realistic}. In these modes, Fock states and SMS states perform worse than the vacuum strategy. Rather it is the coherent state which best discriminates the channels \cite{optimal}. The non-trivial $\mathcal{E}_0$ channel therefore results in a different ordering of the strategies. The behaviour of the SMS state, $\rho_{\text{SMS}}$, is consistent with our earlier analysis: $\mathcal{E}_0(\rho_{\text{SMS}})$ is a squeezed thermal state whose thermal component worsens the discrimination between $\mathcal{I}$ and $\mathcal{E}_{\text{amp}}$ \cite{FockUnkown}. In contrast coherent states remain pure under $\mathcal{E}_0$. Interestingly, the TMS provides near-optimal discrimination implying that entanglement remains a useful resource.

\begin{figure}
\centering
\includegraphics[width=0.45\columnwidth]{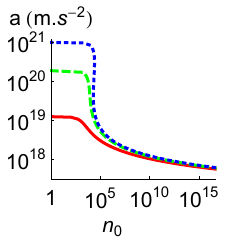}\hspace{.5cm}
\includegraphics[width=0.45\columnwidth]{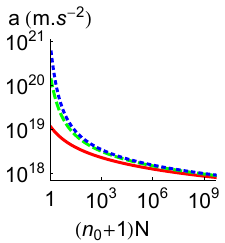}
\caption{\label{fig:accnreduction} (Colour Online) $60$\%  (red thick), $ 95$\% (green dashed) and $99$\% (blue dotted) confidence lines for the discrimination of the Unruh effect as a function of the acceleration and the mean particle number $n_0$. (Left) The practical strategy of a coherent state (displaced in the $x$-direction) in a quasimonochromatic mode at microwave frequencies $10$GHz with a bandwidth of $1$GHz, for a single experimental run; the acceleration can be further lowered by about a factor of one half at the same confidence level by increasing $N$. (Right) The ultimate ideal state: Fock state in a  $10$GHz Unruh mode.}
\end{figure}

The statistical confidence in the discrimination of the two non-inertial channels is plotted in Fig.~\ref{fig:accnreduction} as a function of the acceleration and probe resources ($n_0$ and $N$).  Using physically reasonable energies (resources) we find that reductions of more than three orders of magnitude in the required acceleration for the same level of statistical confidence is possible \cite{redanalogue}.

How large must the acceleration be for the approach to be implemented?  In the case of the actual Unruh effect, if one uses a coherent microwave signal of quasimonochromatic frequency $\omega=10^{10}$Hz containing $10^{10}$ photons, one would be able to discriminate the Unruh theory with a probability of misidentification of approximately (but no less than) $1\%$ by accelerating a Kennedy receiver with an acceleration of $10^{18}$ms$^{-2}$. This should be compared to the acceleration requirement of $10^{21}$ms$^{-2}$ for a photon counting device accelerating through a perfect vacuum state. In analogue settings, as for example described in \cite{Retzker2008} the acceleration required in vacuum can be as low as $a \sim 5\times10^5$ m$s^{-2}$ and thus the quantum statistical tools we have described have the potential to bring these accelerations down to the order of $10^2$m$s^{-2}$.

\section{Discussion}
We have described how an experiment that filters frequencies and performs positive operator valued measurements in the accelerated frame can be used to test the Unruh effect.  We showed, in contrast with the standard theory, that there is a seemingly meaningful way of describing a theory in which an accelerated observer does not detect any particles in the Minkowski vacuum by assigning Unruh modes rather than Rindler modes to the measured frequencies. This is of course related to the discussion about the meaning of the particle concept dating back to the beginnings of the subject of quantum field theory in curved spacetime (see for example the discussion in 3.3 of \cite{Birrell1982}). Perhaps new to this discussion, is that the state of motion of the observer alone may not uniquely define the particle content. Rather what is also important is the mode of operation of the detector. Perhaps there are detector setups in which particles are detected and those in which there are no particles detected. The fact that calculations using an accelerating Unruh-DeWitt detector show a thermal response, is good theoretical justification for the H1 hypothesis, and was largely responsible for the acceptance of this perspective historically. What is still not clear is if there is an accelerated system which operates as a detector according to the H0 hypothesis; perhaps all physical detectors work like Unruh-DeWitt detectors. On the other hand, we note that even the physical realisability of the Unruh-DeWitt detector has been put into question \cite{Narozhny}. Our point of view on the matter is therefore one of impartiality. We have framed the question scientifically in terms of a binary hypothesis test, and devised optimal strategies for ascertaining which of the two hypotheses are realised in any given experimental setup.

Our analysis indicates that the Unruh theory can be tested at lower accelerations using a coherent source at large energies.
It appears to be the most practical strategy and applies even when the modes are quasi-monochromatic with respect to Minkowski time. This is because coherent states remain coherent under the $\mathcal{E}_0$ transformation. While coherent probes behave similarly for both Minkowski-mode and Unruh-mode initial states, in the Minkowski-mode case larger initial intensities are required.   Hence, one would best be able to discriminate the theories if it were possible to engineer initial modes in the Unruh basis.

In parameter estimation \cite{Aspachs,Downes} strategies which best distinguish evolutions $a$ and $a+\delta a$ for some channel parameter $a$ are sought. This differs to channel discrimination, which can be applied (as we have) to discriminate between two independent channels $\mathcal{E}_0$ and $\mathcal{E}_1$. In the case of Unruh modes, we found that Fock states, which were the best states to estimate the temperature in \cite{Aspachs}, also give the best discrimination of the theories. Nevertheless, if one uses realistic quasimonochromatic modes, Fock states are not useful in the discrimination. It would be interesting to know if the same holds for the parameter estimation of temperature.

We have assumed in this work that the detector accelerates uniformly for all of time. Such eternal acceleration ensures that the system is stationary (time independent) and that the detector is on long enough that it will be able to measure a perfect blackbody distribution (right down to the longest wavelengths). In practice, the acceleration can only be for a finite duration. Generally, experimental proposals \cite{ExpProp} consider short constant accelerations or even oscillatory accelerations. Studies suggest \cite{Doukas2013} that even for non-eternal acceleration a particle detector in certain regimes will observe an approximate thermal response (over the frequencies that are measurable during this time interval). Although outside the scope of the current work, it would be interesting to explore how the results presented in this paper would be effected in such non-eternally accelerating situations.

Our analysis can be generalised to any theory with horizons, where the Rindler modes are replaced with modes localised inside or outside of the horizon. Furthermore, the tools of state and channel discrimination are also likely to be of use in other tests of quantum field theory in curved spacetimes especially in analogue experiments \cite{Analogue} where Bogoliubov transformations act.

\appendix
\section{Orthonormal bases for Minkowski and Rindler frames.}
\label{app:orthobases}
In this appendix we give general procedures for constructing non-standard orthonormal bases in the Minkowski and Rindler frames. The motivation for doing this is that we will often want to describe the reduced state of the field in a mode that is not a Minkowski or Rindler plane wave. For this purpose, it is useful to expand the field in a basis for which the mode of interest is a basis function and then trace out the orthogonal subspace.

First we consider the Minkowski frame. Recall that the standard basis functions for solutions to the Klein-Gordon equation in the Minkowski frame are the plane waves $\{u_k, u_k^{\star}\}$. We call the subspace of solutions spanned by the positive $\hat{E}$-frequency Minkowski plane waves, $u_k$, the $E^+$ subspace. Since the $u_k$ have positive norm, it follows that every solution in the positive $E^+$ subspace also has positive norm. Therefore the scalar product (\ref{scalarproduct}) is a genuine inner product on the positive $E^+$ subspace.  In a similar way, one can construct an inner product space for the subspace of solutions spanned by $u_k^{\star}$ using the negative of the scalar product as the (positive definite) inner product. We call this the $E^-$ subspace.

Let $\phi\in E^+$. By Gram-Schmidt orthonormalisation starting with the function $\phi$ one can construct a complete orthonormal basis of functions for the $E^+$ subspace. Label these basis functions $\{\phi^{(i)}\}$ for $i=0,1,2\ldots$ where $\phi^{(0)}=\phi$. A complete orthonormal basis for the $E^-$ subspace is then found by complex conjugation of these functions. These basis functions satisfy the orthonormality relations:
\bea\label{phiorthonormal1}
(\phi^{(i)},\phi^{(j)})&=&\delta_{ij},\\
(\phi^{(i)\star},\phi^{(j)\star})&=&-\delta_{ij},\\
(\phi^{(i)},\phi^{(j)\star})&=&0.\label{phiorthonormal2}
\eea

By associating annihilation (creation) operators, $\hat{A}_i$ ($\hat{A}_i^{\dagger}$), with the positive (negative) norm basis functions, the field operator can be expanded as:
\bea\label{fielddecomp}
\Phi=\sum_i\phi^{(i)}\hat{A}_i+\phi^{(i)\star}\hat{A}_i^{\dagger},
\eea
from which one can identify the relations:
\bea\label{arelations1}
\hat{A}_i&=&(\phi^{(i)},\hat{\Phi}),\\
\hat{A}_i^{\dagger}&=&-(\phi^{(i)\star},\hat{\Phi}). \label{arelations2}
\eea
For the initial mode $\phi$ we will often define $\hat{A}\equiv\hat{A}_0$ without the subscript.

The Bogoliubov transformation corresponding to the change of basis from the standard Minkowski basis into the $\phi$-basis, is found by Fourier decomposing the $\phi^{(i)}$ basis functions in terms of the plane waves:
\bea
\phi^{(i)}=\int dk(u_k,\phi^{(i)})u_k,
\eea
where we have used $(u_k^{\star},\phi^{(i)})=0$. Then using the relations (\ref{arelations1}) and the equivalent relation for the Minkwoski plane waves, i.e., $\hat{a}_k=(u_k,\hat{\Phi})$, one can write:
\bea \label{BogMinkphi}
\hat{A}_i=\int dk (\phi^{(i)},u_k)\hat{a}_k,
\eea

In the Rindler frame the standard basis functions are given by the Rindler modes $\{w_{\text{I}k},w_{\text{II}k}, w_{\text{I}k}^{\star},w_{\text{II}k}^{\star}\}$. In this case both $w_{\text{I}k}$ and  $w_{\text{II}k}^{\star}$ are positive frequency with respect to the boost operator $\hat{K}$.  Let $K_\text{I}^+$ be the subspace spanned by the $w_{\text{I}k}$. Since $w_{\text{I}k}$ have positive norm, the $K_\text{I}^+$ subspace is an inner product space with (\ref{scalarproduct}) as the inner product. Retracing our steps above, if $\psi\in K_\text{I}^+$ then we can find an orthonormal basis $\{\psi^{(i)}\}$  for $K_\text{I}^+$ where $\psi^{(0)}=\psi$, and an orthonormal basis for $K_\text{I}^-$ (the subspace spanned by  $w_{\text{I}k}^{\star}$) by complex conjugation of these functions.

For completeness we mention that the $w_{\text{II}k}^{\star}$ modes are negative norm, so the negative of the scalar product would be a suitable inner product for orthonormalising the space spanned by these functions. However, in this paper we only consider a single observer in the right wedge, so there will not be an occasion in which it is necessary to change the $w_{\text{II}k}^{\star}$ basis.

As before, defining $\hat{d}_i\equiv(\psi^{(i)},\hat{\Phi})$, we find:
\bea
\hat{d}_i=\int dk (\psi^{(i)},w_{\text{I}k})\hat{b}_{\text{I}k}.
\eea
By construction the $\{\psi^{(i)}\}$ are superpositions of $w_{\text{I}k}$ only. This means that they will be mixed superpositions of the positive and negative $\hat{E}$ eigenfunctions. We can therefore write:
\bea
\psi^{(i)}=\int dk(u_k,\psi^{(i)})u_k-(u_k^{\star},\psi^{(i)})u_k^{\star}.
\eea
and find the Bogoliubov transformation for the change of basis from standard Minkowski basis to the $\psi$-basis:
\bea \label{BogMinkpsi}
\hat{d}_i=\int dk (\psi^{(i)}, u_k)\hat{a}_k+(\psi^{(i)},u_k^{\star})\hat{a}_k^{\dagger},
\eea
where we used  $\hat{a}_k=(u_k,\hat{\Phi})$ and $\hat{a}_k^{\dagger}=-(u_k^{\star},\hat{\Phi})$, which follow from the standard decomposition of the field operator similar to equations (\ref{fielddecomp}) and (\ref{arelations1}-\ref{arelations2}).

A few further remarks are in order regarding the validity of (\ref{BogMinkpsi}). We can write the Bogoliubov coefficient in (\ref{BogMinkpsi}) as:
\bea
(\psi_1^{(i)},u_k)&=&\int dk' (w_{\text{I}k'},\psi_1^{(i)})(w_{\text{I}k'}, u_k),\\
                          &=&\frac{i}{2\pi\sqrt{|k|}} \int \frac{dk' }{\sqrt{|k'|}} (w_{\text{I}k'},\psi_1^{(i)}) e^{\pi k'/2a}  (k/a)^{i k'/a}. \label{divint}
\eea
The divergence when $k\rightarrow0$ is of the same kind of divergence as in $u_k$ which arises from the choice of normalisation and is therefore not problematic. On the other hand, because of the factors of $1/\sqrt{|k'|}$ and $e^{\pi k'/2a}$, the integral in equation (\ref{divint}) is potentially both IR and UV divergent. This divergence would lead to the problematic result that the Bogoliubov coefficient was infinite for all $k$. The integral can be insured to be finite if $\psi_1^{(i)}$ is composed of a finite interval of Rindler frequencies of positive $k'$. In this paper we take $\psi_1$ to be a quasimonochromatic mode, that is therefore compactly supported in $\hat{K}$-space. This is physically motivated by the observation that all detectors have a finite spectral bandwidth. A complete orthogonal basis for $K_\text{I}^+$ containing $\psi_1$, that are compactly supported in Rindler frequencies and therefore insured to be well-defined, can be constructed by forming wavepackets in Rindler frequency space. For details of this construction see pg. 18-20 of \cite{Takagi}.

\section{Formal derivation of $\mathcal{E}_0$ and $\mathcal{E}_1$}
\label{app:FormalDerChannels}

In this appendix we find formal expressions for the general $\mathcal{E}_0$ and $\mathcal{E}_1$ channels to complement section \ref{E0E1channels}. First we consider the H0 hypothesis where the detection mode $\psi_0$ is a superposition of positive frequency eigenfunctions of the energy operator, $\hat{E}=i\frac{\partial}{\partial t}$. From (\ref{BogMinkphi}) with $\phi\rightarrow\psi_0$ and we have:
\bea
\hat{A}_i=\int dk (\psi_0^{(i)},u_k)\hat{a}_k.
\eea
Since the operators $\hat{A}_i$ and $\hat{a}_k$ annihilate the Minkowski vacuum state they are unitarily related. The unitary operator, $U_0$, that achieves this change of basis is defined by:
\bea
U_0 \hat{a}_iU_0^{\dagger}\equiv\hat{A}_i=\int dk (\psi_0^{(i)},u_k)\hat{a}_k.
\eea
The $\mathcal{E}_0$ channel is then found by writing the initial state in the $\psi_0$ basis and tracing out the subspace orthogonal to $\psi_0$:
\bea
\mathcal{E}_0(\rho)=\text{Tr}_{\perp{\psi_0}}[U_0\rho U_0^{\dagger}].
\eea
This is the slightly more detailed justification for equation (\ref{E0general}).

Under the H1 hypothesis $\psi_1$ is a superposition of Rindler modes, it will therefore be a mixed superposition of positive frequency and negative frequency eigenfunctions of $\hat{E}$. From (\ref{BogMinkpsi}) with $\psi\rightarrow \psi_1$ we have
\bea \label{dMinkexp}
\hat{d}_i=\int dk (\psi_1^{(i)}, u_k)\hat{a}_k+(\psi_1^{(i)},u_k^{\star})\hat{a}_k^{\dagger}.
\eea
We postulate that there exists a unitary operator, $U_1$, such that:
\bea\label{bogoU1}
U_1\hat{a}_kU_1^{\dagger}\equiv\hat{d}_i=\int dk (\psi_1^{(i)},u_k)\hat{a}_k+(\psi_1^{(i)},u_k^{\star})\hat{a}_k^{\dagger}.
\eea

We can then write $\mathcal{E}_1$ as:
\bea\label{E1problematic}
\mathcal{E}_1(\rho)=\text{Tr}_{\perp \psi_1}[U_1\rho U_1^{\dagger}].
\eea
Thus, reproducing equation (\ref{E1general}). However, there is no guarantee that a Unitary operator relating a state in the Minkowski frame with a state in the Rindler frame exists.  In fact, the Minkowski and Rindler vacua (\ref{RindlerToMink}) are Unitarily inequivalent, see for example the discussion on pg. 31 of \cite{Takagi}. This is ordinarily dealt with by working ``mode-by-mode.''

The problem with (\ref{E1problematic}) is that it may not be possible to perform a Unitary operation on the state $\rho\rightarrow U_1\rho U_1^{\dagger}$. The key is to work with the operators themselves rather than the states. Operators on the $\psi_1$ subspace are easily expressed in the Minkowski plane-wave basis using the Bogoliubov transformation (\ref{bogoU1}). Therefore, all expectation values of quantities measured on the $\psi_1$ subspace can be calculated by writing the operator in the Minkowski plane wave basis. For example, when the state is a Gaussian state it is completely characterised by its first and second moments. These are simply expectation values of operators defined on the $\psi_1$ subspace. Therefore, the state on the $\psi_1$ subspace can be completely determined even though the vacua may not be Unitarily related. Further details on this derivation of the channel in the Gaussian case are provided in section \ref{sec:XYderive}.

\section{Derivation of the general mode Unruh channel for Gaussian states}\label{sec:XYderive}

In this appendix we will describe how the covariance matrix formalism can be used to find the channel acting on Gaussian states. In particular we derive the transformation matrices found in equation (\ref{H0GausTransfs}). These matrices completely categorise the Unruh channel on Gaussian states for general single mode preparations and single mode measurements. We first consider the action of the Unruh channel on a general mode coherent state. Not only are coherent states relevant to Section \ref{displaced} but quite remarkably the information we gain from investigating the coherent state is enough to deduce the general form of the channel matrices for any Gaussian state.

The technique follows that described in \cite{Dragan2012}. Consider a state of the field that is almost entirely (Minkowski) vacuum except for a single mode that is populated in the form of a coherent state, $|\alpha\rangle$. The populated mode could be a plane wave, an Unruh mode, a Gaussian wavepacket, or any other mode shape of interest. Assume that the positive norm solution associated with this mode, $\phi$, is a superposition of purely positive frequencies with respect to $\hat{E}$.  We can then find a complete orthonormal basis of functions $\{\phi^{(i)}\}$ for the positive $\hat{E}$ subspace with $\phi^{(0)}=\phi$, see Appendix \ref{app:orthobases}.

Next consider a general measurement of the field which may occur in a mode that is different to the one in which the field was prepared. For example, one might prepare a broadband wavepacket mode but then only select out and measure the state of a narrow band of frequencies from this original source using filters and other devices. Label the measured mode by $\psi$.  In fact, for the sake of generality, we will assume that $\psi$ is not necessarily in the positive $\hat{E}$ subspace. It can then be written as a superposition of both positive and negative $\hat{E}$ eigenfunctions.  Using the relations (\ref{phiorthonormal1}-\ref{phiorthonormal2}) we can write:
\bea\label{psidecomp}
\psi=\sum_i(\phi^{(i)},\psi)\phi^{(i)}-(\phi^{(i)\star},\psi)\phi^{(i)\star}.
\eea
Defining the annihilation operator $\hat{d}=(\psi,\hat{\Phi})$, and using (\ref{arelations1}-\ref{arelations2}) we obtain the operator decomposition \cite{convention}:
\bea\label{ddecomp}
\hat{d}=\sum_i(\psi,\phi^{(i)})\hat{A}_i+(\psi,\phi^{(i)\star})\hat{A}_i^{\dagger}.
\eea
For brevity, we separate the $i=0$ terms from the sum, and define a new operator $\hat{d}'$ equal to the remaining terms:
\bea
\hat{d}'\equiv\sum_{i\neq0}(\psi,\phi^{(i)})\hat{A}_i+(\psi,\phi^{(i)\star})\hat{A}_i^{\dagger},
\eea
so that (\ref{ddecomp}) becomes:
\bea\label{drelation}
\hat{d}=(\psi,\phi)\hat{A}+(\psi,\phi^{\star})\hat{A}^{\dagger} +\hat{d}'.
\eea

The $\hat{d}$ operator acts of the subspace that describes those excitations of the field accessible to our detector (alternatively, it can be thought of as an operator on the detector subspace itself).

We first define the $\hat{x}$ and $\hat{p}$ quadrature operators by: $\hat{x}\equiv(\hat{d}+\hat{d}^{\dagger})$ and $\hat{p}\equiv\frac{1}{i}(\hat{d}-\hat{d}^{\dagger})$. Arranging these elements into a column vector $\hat{\mathbf{R}}=(\hat{x},\hat{p})^T$, we can define the mean value (also known as the first moment):
\bea \label{mean}
\mathbf{\overline{x}}\equiv \langle\hat{\mathbf{R}}\rangle
\eea
and the covariance matrix (also known as the second moment):
\bea\label{covariance}
\overline{V}_{ij}= \frac{1}{2}\langle \hat{\mathbf{R}}_i\hat{\mathbf{R}}_j+\hat{\mathbf{R}}_j\mathbf{R}_i\rangle-\langle \hat{\mathbf{R}}_i\rangle\langle \hat{\mathbf{R}}_j\rangle,
\eea
where the expectation values, $\langle\cdot\rangle$, are taken with respect to the initial state (assumed here to be a coherent state). Note that the vacuum is normalised such that it's covariance matrix is the identity.

Gaussian states are defined as those states whose Wigner function is Gaussian \cite{Holevo75} (see pg. 5-6 of \cite{Weedbrook2012} for a review).  They are completely characterised by their first and second moments only. A Gaussian state remains Gaussian if it undergoes a Gaussian transformation. Since linear Bogoliubov transformations and trace operations are Gaussian operations, changing basis from Minkowski to Rindler frames is a Gaussian transformation.

Equations (\ref{mean}) and (\ref{covariance}) can be written in the expanded form:
\bea
\mathbf{\overline{x}}= (\langle\hat{x}\rangle~\langle\hat{p}\rangle )^T,
\eea
and
\bea
\overline{V}= \left(\begin{array}{c c}
                    \langle \hat{x}^2\rangle-\langle \hat{x}\rangle^2 & \frac{1}{2}\langle \hat{x} \hat{p}+\hat{p}\hat{x} \rangle-\langle\hat{x}\rangle\langle \hat{p}\rangle\\
                    \frac{1}{2}\langle \hat{x} \hat{p}+\hat{p}\hat{x} \rangle-\langle\hat{x}\rangle\langle \hat{p}\rangle
& \langle \hat{p}^2\rangle-\langle \hat{p}\rangle^2
                   \end{array}\right).
                    \eea
The initial coherent state can be written in terms of a displacement operator of the $\phi$ mode, acting on the (Minkowski) vacuum state, $D_{\phi}(\alpha)|0\rangle$.

When acting on annihilation operators of the same mode, the displacement operator satisfies the relation:
\bea
D_{\phi}(\alpha)^{\dagger}\hat{A}D_{\phi}(\alpha)=\hat{A}+\alpha.
\eea
On the other hand, the displacement operator passes straight through operators, like $\hat{d}'$, that commute with $\hat{A}$:
\bea
D_{\phi}(\alpha)^{\dagger}\hat{d}'D_{\phi}(\alpha)=\hat{d}',
\eea
where we have also used the unitarity of the displacement operators to obtain the r.h.s.

Thus, using equation (\ref{drelation}) we obtain:
\bea
D_{\phi}(\alpha)^{\dagger}\hat{d}D_{\phi}(\alpha)=\hat{d}'+(\psi,\phi)(\hat{A}+\alpha)+(\psi,\phi^{\star})(\hat{A}^{\dagger} +\alpha^{\star}).
\eea
Now again using equation (\ref{drelation}) to eliminate the $\hat{d}'$ operator from the r.h.s we obtain:
\bea
D_{\phi}(\alpha)^{\dagger}\hat{d}D_{\phi}(\alpha)=\hat{d}+(\psi,\phi)\alpha+(\psi,\phi^{\star})\alpha^{\star}.
\eea
And consequently,
\bea
D_{\phi}(\alpha)^{\dagger}\hat{x}D_{\phi}(\alpha)&=&\hat{x}+2\text{Re}\left[(\psi,\phi)\alpha+(\psi,\phi^{\star})\alpha^{\star}\right],\\
D_{\phi}(\alpha)^{\dagger}\hat{p}D_{\phi}(\alpha)&=&\hat{p}+2\text{Im}\left[(\psi,\phi)\alpha+(\psi,\phi^{\star})\alpha^{\star}\right].
\eea
Next we notice from (\ref{ddecomp}) that expectation of $\hat{d}$ in the Minkowski vacuum state vanishes. This is because the $\hat{A}_i$ operators annihilate the Minkowski vacuum state, and the $\hat{d}$ operator is a linear superposition of such operators and their conjugates. Similarly, the expectation value of the $\hat{x}$ and $\hat{p}$ operators also vanish when taken with respect to the Minkowski vacuum state.  Therefore, the first moments are given by:

\bea
\mathbf{\overline{x}}'&=& \left(\!\!\begin{array}{c}2\text{Re}\left[(\psi,\phi)\alpha+(\psi,\phi^{\star})\alpha^{\star}\right]\\2\text{Im}\left[(\psi,\phi)\alpha+(\psi,\phi^{\star})\alpha^{\star}\right]\end{array}\!\!\right)\\
  &=& \left(\!\!\begin{array}{cc}
\text{Re}[(\psi_1,\phi)+(\psi_1,\phi^{\star})]&-\text{Im}[(\psi_1,\phi)+(\psi_1,\phi^{\star})] \\
\text{Im}[(\psi_1,\phi)-(\psi_1,\phi^{\star})] & \text{Re}[(\psi_1,\phi)-(\psi_1,\phi^{\star})]
\end{array}\!\!\right) \left(\!\!\begin{array}{c} 2 \text{Re} (\alpha)\\2 \text{Im} (\alpha)\end{array}\!\!\right). \nonumber \\&&
\eea
On the last line we have re-expressed the moments in terms of a product of a matrix (that is independent of $\alpha$) and a column vector. But the column vector, $\mathbf{\overline{x}}=(2 \text{Re} (\alpha) ~ 2 \text{Im} (\alpha))^T$, is nothing other than the first moment of the coherent state in the $\phi$ basis. In general, non-displacing Gaussian channels transform the first moments according to $\overline{\mathbf{x}}'=X\overline{\mathbf{x}}$. Since the state $\alpha$ was arbitrary, the transformation matrix, $X$, must therefore be:
\bea
X=\left(\begin{array}{cc}
\text{Re}[(\psi_1,\phi)+(\psi_1,\phi^{\star})]&-\text{Im}[(\psi_1,\phi)+(\psi_1,\phi^{\star})] \\
\text{Im}[(\psi_1,\phi)-(\psi_1,\phi^{\star})] & \text{Re}[(\psi_1,\phi)-(\psi_1,\phi^{\star})]
\end{array}\right). \nonumber \\
\eea

We next calculate the covariance matrix. In order to do so we need to calculate the terms $\langle \hat{x}^2\rangle$, $\langle \hat{p}^2\rangle$ and $\langle \hat{x}\hat{p}\rangle$. Since the calculations are similar for each case we will only demonstrate the method for $\langle \hat{x}^2\rangle$ and provide the results for the others at the end.

\bea
\langle \alpha| \hat{x}^2|\alpha\rangle&=&\langle 0|(D^{\dagger}(\alpha)\hat{x}D(\alpha))^2|0\rangle\\
 &=&\langle 0|\left(\hat{x}+2\text{Re}\left[(\psi,\phi)\alpha+(\psi,\phi^{\star})\alpha^{\star}\right]\right)^2\!|0\rangle\\
 &=&\langle 0|\hat{x}^2|0\rangle+4\text{Re}\left[(\psi,\phi)\alpha+(\psi,\phi^{\star})\alpha^{\star}\right]^2.
\eea

The first term can be calculated by writing $\hat{d}$ as:
\bea
\hat{d}=\int dk (\psi, w_{\text{I}k}) \hat{b}_{\text{l}k}
\eea
and using
\bea
\hat{b}_{\text{l}k} = \cosh{r_k}\hat{A}_{\text{R}k}+\sinh{r_k}\hat{A}_{\text{L}k}^{\dagger},
\eea
which follows from equations (\ref{UnruhRindlerOps1}-\ref{UnruhRindlerOps2}). We find:
\begin{equation}
\langle 0|(\hat{d}+\hat{d}^{\dagger})^2|0\rangle=1+2 n,
\end{equation}
where $n$ is given by equation (\ref{noise}). It then follows that:
\bea
\overline{V}_{11}'&=&\langle \hat{x}^2\rangle-\langle \hat{x}\rangle^2=2n +1
\eea
The other elements are found a similar way. In summary, we obtain, $\overline{V}'=(2n +1)\mathbb{I}_{2}$, where $\mathbb{I}_{2}$ is the $2\times 2$ identity matrix.

In the $\phi$ basis the covariance matrix of the initial state is just the identity: $\overline{V}=\mathbb{I}_{2}$ (i.e., a coherent state). Furthermore, since the general form for a single mode Bosonic Gaussian channel can be written as:
\bea
\overline{V}'=X\overline{V}X^{T}+Y,
\eea
we can deduce that the matrix $Y$ must take the form:
\bea
Y&=&\overline{V}'-X\overline{V}X^{T}\\
&=&(2n +1)\mathbb{I}_{2}-XX^{T},
\eea
as was to be shown. To obtain the results under H0 set $\psi=\psi_0$ and $n=0$. For H1, set $\psi=\psi_1$.

\section{Derivation of the Unruh channel for Fock states}\label{app:Fock}

We present here the derivation of the density matrices in equations (\ref{eqn:Fock1}-\ref{eqn:Fock2}) for Fock states prepared in a single mode, $\phi$, and measured in another mode $\psi$, where $\psi$ is taken to be a quasimonochromatic mode.

Consider first two inertially defined positive frequency modes, $\phi$, and $\psi$ (this corresponds to the assumptions of the H0 hypothesis). Since they are vectors in the usual Hilbert space of positive frequency solutions, we can decompose $\phi$ into a part parallel and a part orthogonal to $\psi$:
\begin{equation}
\phi=(\psi_{\perp},\phi)\psi_{\perp}+(\psi,\phi)\psi.
\end{equation}
Unit normalisation for each of the modes allows us to write $(\psi_{\perp},\phi)=e^{i\theta}\sqrt{1-|(\psi,\phi)|^2}$, where $\theta$ is some unknown phase. The annihilation operators associated with these modes are, $\hat{A}=(\phi,\hat{\Phi})$, $\hat{d}_{\perp}=(\psi_{\perp},\hat{\Phi})$ and $\hat{d}=(\psi,\hat{\Phi})$. Then:
\begin{equation}
\hat{A}=e^{-i\theta}\sqrt{1-|(\psi,\phi)|^2} \hat{d}_{\perp}+(\psi,\phi)^{\star} \hat{d},
\end{equation}
and using the binomial theorem:
\begin{eqnarray}\label{an}
\hat{A}^N=\sum_{k=0}^N
\left(\!\! \begin{array}{c}N\\ k
\end{array}\!\!\right)
\left(e^{-i\theta}\sqrt{1-|(\psi,\phi)|^2}\right)^{N-k}\!\!\!\! (\psi,\phi)^{\star k} \hat{d}_{\perp}^{N-k}\hat{d}^{k}.
\end{eqnarray}
The density matrix of an $n_0$ particle Fock state in the mode $\phi$ measured in the mode $\psi$ is then:
\begin{equation}
\rho_0=\text{Tr}_{\psi_{\perp}}|(n_0)_{\phi}\rangle\langle (n_0)_{\phi} |,
\end{equation}
where $|(n_0)_{\phi}\rangle\equiv\frac{\hat{A}^{\dagger n_0}}{\sqrt{n_0!}}|0\rangle$. Using (\ref{an}) and taking the trace inside the summation one obtains:

\begin{eqnarray}
\rho_0=\sum_{k=0}^{n_0}
\left( \begin{array}{c}n_0\\ k
\end{array}\right)
\big|1-|(\psi,\phi)|^2\big|^{n_0-k} |(\psi,\phi)|^{2 k} |k_{\psi}\rangle\langle k_{\psi} |. \label{eqn:FockUnruhmeasurement}
\end{eqnarray}
With $\psi=\psi_0$ this state corresponds to the part of the initial Fock state state that is accessible to the detector (or if you prefer, to the state of detector state itself) under the H0 hypothesis, cf equation (\ref{eqn:Fock1}).

We now consider the specific case when $\psi$ is quasi-monochromatic about a Rindler mode, $\Omega$. We first transform into the Unruh mode basis by putting $\psi=\psi_0$ into equation (\ref{eqn:FockUnruhmeasurement}), then we use the transformation found in \cite{Aspachs} (above equation (3))  to transform the Unruh basis into the Rindler basis:
\begin{eqnarray}
\rho_1=\sum_{k=0}^{n_0}\sum_{i=0}^{\infty}
\left( \begin{array}{c}{n_0}\\ k
\end{array}\right)
\big|1-|(\psi_0,\phi)|^2\big|^{{n_0}-k} |(\psi_0,\phi)|^{2 k} \\ \nonumber \times \ C_{k,i}(\Omega) |(k+i)_{\psi}\rangle\langle (k+i)_{\psi} |,
\end{eqnarray}
where $C_{k,i}(\Omega)=\left( \begin{array}{c}k+i\\ k
\end{array}\right) (\cosh r_{\Omega})^{-2(k+1)}\tanh^{2i}{ r_{\Omega}} $.
Noting that $n=\sinh^2{r_{\Omega}}$ and $r_{\Omega}={\rm arctanh}(e^{-\pi \Omega})$ this then completes the derivation of equation (\ref{eqn:Fock2}).

\section{Numerical implementation of the realistic modes}\label{app:numeric}
In this paper the quasimonochromatic wavepacket scalar products are numerically calculated by performing several double integrals. In this appendix we provide more details on how these double integrals are calculated.

As described in the main text we assume that the modes are quasimonochromatic. Let the source mode have a central wavenumber $k_\text{M}$ and spectral width $\Delta k_\text{M}$ and let the detectors operate at either an Unruh wavenumber $k_\text{U}$ or Rindler wavenumber $k_\text{I}$ with spectral widths of $\Delta k_\text{U}$ and $\Delta k_\text{I}$ respectively. Assuming the quasimonochromatic modes to be a uniform box of wavenumbers we can write:
\bea
\phi&=&\frac{1}{\sqrt{\Delta k_\text{M}}}\int_{k_\text{M}}^{k_\text{M}+\Delta k_\text{M}}u_k dk,\\
\psi_1&=&\frac{1}{\sqrt{\Delta k_\text{I}}}\int_{k_\text{I}}^{k_\text{I}+\Delta k_\text{I}}w_{\text{I}k} dk,\\
\psi_0&=&\frac{1}{\sqrt{\Delta k_\text{U}}}\int_{k_\text{U}}^{k_\text{U}+\Delta k_\text{U}}u_{\text{R}k} dk,
\eea
where $u_k$ are Minkowski plane waves, $w_{\text{I}k}$ are right-wedge Rindler plane waves, and $u_{\text{R}k}$ are Right-Unruh modes. We therefore have,
\bea
\!\!\!\!\!\!\!\!(\psi_1,\phi)&=&\frac{1}{\sqrt{\Delta k_\text{M}\Delta k_\text{I}}}\int_{k_\text{I}}^{k_\text{I}+\Delta k_\text{I}}\!\!\int_{k_\text{M}}^{k_\text{M}+\Delta k_\text{M}}(w_{\text{I}k} ,u_{k'})dk dk'\label{eqn:numInphi},\\
\!\!\!\!\!\!\!\!(\psi_1,\phi^{\star})&=&\frac{1}{\sqrt{\Delta k_\text{M}\Delta k_\text{I}}}\int_{k_\text{I}}^{k_\text{I}+\Delta k_\text{I}}\!\!\int_{k_\text{M}}^{k_\text{M}+\Delta k_\text{M}}(w_{\text{I}k} ,u_{k'}^{\star})dk dk'\label{eqn:numInphistar},\\
\!\!\!\!\!\!\!\!(\psi_0,\phi)&=&\frac{1}{\sqrt{\Delta k_\text{M}\Delta k_\text{U}}}\int_{k_\text{U}}^{k_\text{U}+\Delta k_\text{U}}\!\!\int_{k_\text{M}}^{k_\text{M}+\Delta k_\text{M}}(u_{\text{R}k} ,u_{k'})dk dk'\label{eqn:numInphizero}.
\eea
The inner products $(w_{\text{I}k} ,u_{k'})$ and $(w_{\text{I}k} ,u_{k'}^{\star})$ can be calculated explicitly (using the contour trick in \cite{Takagi} see pg. 24):
\bea
(w_{\text{I}k} ,u_{k'})&=&\frac{i}{2\pi} \frac{e^{\pi k/2a}}{\sqrt{|k k'|}} (k'/a)^{i k/a}\label{KGwIu},\\
(w_{\text{I}k} ,u_{k'}^{\star})&=&\frac{i}{2\pi} \frac{e^{-\pi k/2a}}{\sqrt{|k k'|}} (k'/a)^{i k/a}.
\eea
\begin{widetext}
With these explicit expressions for the integrands the integrals (\ref{eqn:numInphi}) and (\ref{eqn:numInphistar}) can be done numerically. To do the last integral we need to calculate $(u_{\text{R}k} ,u_{k'})$, which we can do using the equations (\ref{UnruhRindler1}-\ref{UnruhRindler2}). From the first we get:
\bea
(\psi_0,\phi)&=&\frac{1}{\sqrt{\Delta k_\text{M}\Delta k_\text{U}}}\int_{k_\text{U}}^{k_\text{U}+\Delta k_\text{U}}\int_{k_\text{M}}^{k_\text{M}+\Delta k_\text{M}}\cosh r_k (w_{\text{I}k},u_{k'})+\sinh r_k (w_{\text{II}k}^{\star} ,u_{k'})dk dk',
\eea
and from the second, which we note is negative frequency w.r.t Minkowski time, and therefore $(u_{\text{L}k}^{\star},u_{k'})=0$, we obtain:
\bea
 (w_{\text{II}k}^{\star} ,u_{k'})=-\tanh r_k (w_{\text{I}k},u_{k'})
\eea
Thus,
\bea
(\psi_0,\phi)&=&\frac{1}{\sqrt{\Delta k_\text{M}\Delta k_\text{U}}}\int_{k_\text{U}}^{k_\text{U}+\Delta k_\text{U}}\int_{k_\text{M}}^{k_\text{M}+\Delta k_\text{M}}(\cosh r_k -\sinh r_k\tanh r_k )(w_{\text{I}k},u_{k'})dk dk'
\eea
or
\bea
(\psi_0,\phi)&=&\frac{1}{\sqrt{\Delta k_\text{M}\Delta k_\text{U}}}\int_{k_\text{U}}^{k_\text{U}+\Delta k_\text{U}}\int_{k_\text{M}}^{k_\text{M}+\Delta k_\text{M}}\sqrt{1-\exp({-2\pi |k|/a})}(w_{\text{I}k},u_{k'})dk dk'.
\eea
These integrals can then be computed numerically by again making use of equation (\ref{KGwIu}).
In the text we impose the conditions, $k_\text{M}=\omega$, $\Delta k_\text{M}=\delta\omega$ and $k_\text{I}=k_\text{U}=\omega_\text{R}$ and $\Delta k_\text{U}=\Delta k_\text{I}=\delta\omega_\text{R}$  (note we work in units with $c=1$).
\end{widetext}

\acknowledgments
We thank I. Fuentes for useful discussions and R. Demkowicz-Dobrzanski for providing suggestions and corrections to the manuscript.   S.P. acknowledges support from EPSRC and the Leverhulme Trust. GA thanks the Tsinghua-Nottingham Teaching and Research Fund for financial support. AD is funded by National Science Center, Sonata BIS grant 2012/07/E/ST2/01402.


\begin{thebibliography}{99}
\bibitem{UnitaryInequiv}
Strictly speaking this is not a unitary transformation, since the vacua are not unitarily related. However, this is only a superficial problem that can be rectified by interpreting the relation on a mode-by-mode basis, for further discussion see pg. 31 of \cite{Takagi}.
\bibitem{footNull}
`Null' is used here to convey both the statistical sense of being a null hypothesis and the fact that the vacuum in the accelerated frame is void of particles in this theory.
\bibitem{convention}
Note we take the convention that the Klein-Gordon indefinite scalar product is anti-linear in the first argument.
\bibitem{uLjustify}
We will consider an accelerated observer in the right wedge who makes measurements on wavepackets that are predominantly localised in this wedge. Because the modulus of $u_{\text{L}k}$ is negligible in this wedge, in our analysis of the H0 hypothesis we will for simplicity only consider contributions from Right-Unruh modes in the detection mode wavepacket.
\bibitem{symbolconvention}
We use the convention that positive norm solutions are denoted by a symbol (i.e., $u$, $w$ etc), so the negative norm solutions are denoted explicitly by the complex conjugation of those symbols.
\bibitem{beta0}
Note that $(\psi_0,\phi^{\star})$ is zero in this case.
\bibitem{footFock}
In fact, since the number of particles in $\mathcal{E}_1(\rho)$ is larger than $n_0$, the $E_0$ strategy of detecting $n_0$ particles or less would also give an equivalent result, although in the ideal case, one should never expect to measure less than $n_0$ particles.
\bibitem{optimal}
The coherent state numerically outperformed $10000$ random Gaussian states at fixed energy suggesting that it is the optimal single mode Gaussian state.
\end{thebibliography}
\end{document}